\documentclass{ifacconf}

\usepackage{balance}
\usepackage{amsmath,amssymb,amsfonts,mathrsfs}
\usepackage{float}
\usepackage{algorithm}
\usepackage{algcompatible}
\usepackage{balance}
\usepackage{textcomp}
\usepackage{subcaption}
\usepackage{color}
\usepackage{graphicx}      
\usepackage{caption}
\usepackage{natbib}        
\newtheorem{remark}{Remark}

\newtheorem{assumption}{Assumption}

\newtheorem{theorem}{Theorem}
\newtheorem{lemma}{Lemma}

\newtheorem{definition}{Definition}

\begin{document}
\begin{frontmatter}

\title{\makebox[\textwidth][l]{\hspace*{-1.7cm} Can Inherent Communication Noise Guarantee Privacy}\\ in Distributed Cooperative Control\,\textcolor{black}{?}\thanksref{footnoteinfo}} 

\thanks[footnoteinfo]{This work was supported by Engineering and Physical Sciences Research Council (EPSRC) of UK Research and Innovation (UKRI) under Grant EP/W524335/1.}

\author[First]{Yuwen Ma}
\author[First]{ Sarah K. Spurgeon}
\author[Second]{Tao Li} 
\author[First]{Boli Chen} 
\address[First]{Electronic and Electrical Engineering Department, University College London, London, Unite Kingdom (e-mail:\\ \{{yuwen.ma.24}, {s.spurgeon}, {boli.chen}\}@ucl.ac.uk).}
\address[Second]{Academy of Mathematics and Systems Science, Chinese Academy of Sciences, Beijing, China (e-mail: litao@amss.ac.cn).}

\begin{abstract}                
This paper investigates privacy-preserving distributed cooperative control for multi-agent systems within the framework of differential privacy. In cooperative control, communication noise is inevitable and is usually regarded as a disturbance that impairs coordination. This work revisits such noise as a potential privacy-enhancing factor. A linear quadratic regulator (LQR)-based framework is proposed for agents communicating over noisy channels, \textcolor{black}{where the noise variance depends on the relative state differences between neighbouring agents.} The resulting controller achieves formation while protecting the reference signals from inference attacks. It is analytically proven that the inherent communication noise can guarantee bounded $(\epsilon,\delta)$-differential privacy without adding dedicated privacy noise, while the \textcolor{black}{system cooperative tracking error} remains bounded and convergent in both the mean-square and almost-sure sense. 
\end{abstract}


\end{frontmatter}

\section{Introduction}
In recent years, distributed cooperative control has become a highly active research area owing to its broad applications in mobile \textcolor{black}{robotics}, unmanned aerial
vehicles (UAVs), and social networks \citep{li2018cooperative,cao2025proximal,d2025hierarchical}. A significant \textcolor{black}{number} of research results \textcolor{black}{have been} reported \citep{oh2015survey,wang2022cooperative}. The primary objective of distributed cooperative control is to drive a group of autonomous agents to achieve common goals or tasks through information exchange among neighbouring agents. However, the data shared among agents may give rise to potential privacy leakage, as external eavesdroppers can extract sensitive information therein, such as the reference signal, control preferences, and individual system dynamics. In fact, \textcolor{black}{it has been} proven that in the absence of privacy-preserving mechanisms, an eavesdropper can reconstruct the training data from shared state information \citep{wang2023tailoring}.

To address privacy concerns, differential privacy (DP) provides a rigorous framework to quantify and manage the trade-off between privacy and performance \citep{dwork2006calibrating,dwork2014algorithmic}. Compared with other privacy-preserving techniques, DP is attractive as it is immune to post-processing, robust against side information, and provides a quantifiable and tunable level of privacy protection. Moreover, as systems evolve toward tighter human–machine integration within cyber-physical-human systems \citep{annaswamy2023cyber}, DP offers a principled mechanism that enables users to interactively adjust system behavior while maintaining formal privacy guarantees.

Up to now, a variety of differentially private distributed cooperative control frameworks have been investigated \citep{huang2012differentially,cortes2016differential,nozari2015differentially,nozari2017differentially,hawkins2020differentially}. In \cite{nozari2017differentially}, the authors introduced a differentially private consensus mechanism to protect agents’ initial states and established a trade-off between DP and consensus error. They proved that, within their algorithmic framework, it is impossible to achieve both accurate consensus and bounded differential privacy simultaneously. This fundamental dilemma has also been reported in \cite{cortes2016differential,nozari2015differentially, huang2024differential}. \cite{nozari2015differentially} discussed the optimal selection of noise parameters for a differentially private average consensus algorithm. However, to the best of our knowledge, the above differentially private distributed cooperative designs, as well as other existing results, rely entirely on actively injected noise to achieve privacy. The impact of inherent communication noise has largely been overlooked. In fact, a similar insight can be found in the wireless communication field, where inherent communication noise (modeled as Gaussian white noise) has been leveraged to enhance privacy in federated learning frameworks \citep{liu2020privacy,liu2022wireless,liu2024differentially}. Notably, passive factor such as stochastic quantizers has recently been employed to achieve differential privacy in single-system tracking control \citep{liu2026design}. Nevertheless, due to the structure of the designed algorithms, bounded DP cannot be guaranteed solely through communication noise \citep{liu2022wireless,liu2024differentially}. Although \cite{liu2020privacy} shows that communication noise alone can ensure privacy in federated learning, this conclusion holds only when the gradient of the cost function is bounded in relation to the local dataset size, the communication channel parameters and privacy budget. These insights motivate a fundamental question: can inherent communication noise itself, when properly integrated into a distributed cooperative control mechanism, guarantee bounded privacy without the need for any actively injected noise?

 In this work, the cooperative objective is to achieve \textcolor{black}{certain agreement}, including scenarios such as formation consensus in multi-robot systems and phase synchronization in power grid networks. A differentially private distributed linear quadratic regulator (LQR) algorithm is proposed, which leverages inherent communication noise to ensure differential privacy of the individual reference signals. It is worth noting that the proposed approach guarantees bounded $(\epsilon,\delta)$-differential privacy and bounded formation tracking simultaneously, even as the number of iterations tends to infinity. This demonstrates that, unlike conventional differentially private control methods that actively inject artificial noise for privacy protection \citep{huang2012differentially,hawkins2020differentially}, the proposed method utilizes existing communication noise, thereby achieving privacy for free without sacrificing additional cooperative control performance.

 \textbf{Notation:} ${\mathbb{R}^{m\times n}}$ and ${\mathbb{R}^{n}}$ are the sets of all ${m\times n}$-dimensional real matrices and $n$-dimensional real vectors, respectively. The sets $\mathbb{R}^{n\times n}_{>0}$ and $\mathbb{R}^{n\times n}_{\ge 0}$ represent positive definite and positive semi-definite real matrices. Complex number and vector spaces are written as $\mathbb{S}$ and $\mathbb{S}^n$, respectively. $\mathbb{N}$ denotes the set of natural numbers, while $\mathbb{N}_+$ stands for the set of positive integers. ${\mathbf{I}}$ is an identity matrix with appropriate dimensions, $\mathbf{0}$ is a zero matrix, and ${\mathbf{1}}$ is a column vector of ones. $\mathrm{diag}\left\{ {{A}_{1}},\ldots, {{A}_{n}}\right\}$ is a diagonal matrix with ${{{A}_{1}},\ldots,{{A}_{n}}}$ as the diagonal elements. {$\mathrm{range}(\cdot)$} indicates the range of an operator. ${\left\| \cdot \right\|_\infty}$ stands for the $\ell_\infty$-norm. Unless otherwise specified, $\|\cdot\|$ represents the \( \ell_2 \)-norm. The space $\ell_2$ is defined as the set of all real sequences whose squared magnitudes are summable, i.e., $\ell_2 = \{ x=(x_n)_{n=1}^{\infty} \mid \sum_{n=1}^{\infty} |x_n|^2 < \infty \}$. $\otimes $ denotes the Kronecker product. Given a square matrix $A\in \mathbb{R}^{n\times n}$, $\lambda_{0}(A)$ denotes the smallest positive eigenvalue, and {$\mathrm{tr}(A)$ represents its trace.} For a symmetric matrix $A$, the notation $A \succ 0$ and $A \succeq 0$ indicate that $A$ is positive definite and positive semi-definite, respectively. The mathematical expectation is denoted by $\mathbb{E}[\cdot]$. $\mathbb{P}(B)$ is the probability of event $B$. Finally, $\mathcal{N}(0,\,\Sigma)$ denotes a zero-mean Gaussian distribution with covariance matrix $\Sigma$. For two sequences $a(t)$ and $b(t)$, we write $a(t) = O(b(t))$ if there exists a constant $c > 0$ and a time $t_0$ such that $|a(t)| \le c |b(t)|,\, \forall\, t \ge t_0$.
 
\section{Preliminaries and Problem Statement} 
\subsection{Multi-agent Systems}
Consider a multi-agent system (MAS) of $N$ agents with a discrete time single integrator model:
\begin{equation}\label{e1}
    x_i(t+1)=x_i(t)+u_i(t),\ i=1,2,\ldots,N,
\end{equation}
where $u_i(t)\in \mathbb{R}^n$ and $x_i(t)\in \mathbb{R}^n$ represent the input and state (position), respectively. The initial state $x_i(0)$ is deterministic for all $i=1,2,\ldots,N.$

The MAS is connected through an undirected communication graph denoted by $\mathcal{G}(\mathcal{V} ,\mathcal{E},\mathcal{A} )$. The graph $\mathcal{G}$ is characterized by a node set $\mathcal{V}={1,2,\ldots,N}$, an edge set $\mathcal{E}\subseteq \mathcal{V}\times\mathcal{V}$ and an adjacency matrix $\mathcal{A}=\left[\begin{matrix}a_{ij}\end{matrix}\right]\in \mathbb{R}^{N\times N}$. Let $N_\mathcal{E}=|\mathcal{E}|$ denote the number of edges and label the edges as $l_1,l_2,\ldots,l_{N_\mathcal{E}}$ following the node index order. For $(j,i)\in\mathcal{E}$, there exists a communication channel between node $j$ and node $i$, indicating that node $j$ is a neighbour of node $i$. The set of \textcolor{black}{neighbours} of node $i$ is denoted by $\mathcal{N}_i=\{j\in\mathcal{V}\mid (j,i)\in\mathcal{E}\}$. Moreover, if $j\in {\mathcal{N}}_{i}$, then $a_{ij}=1$ and $a_{ii}=0$ otherwise. The edge weight matrix $W$ is then defined as $W\triangleq\mathrm{diag}\{a_{ij}\}=\mathbf{I}_{{N_\mathcal{E}}}$, $(i,j)=l_1,\ldots,l_{N_\mathcal{E}}$. An orientation
on $\mathcal{G}$ is the assignment of a direction to each edge. Then, an incidence matrix $B=[b_{ij}]\in\mathbb{R}^{N_\mathcal{E}\times N}$ of $\mathcal{G}$ is defined as $b_{ij}=1$ if edge $i$ enters node $j$, $b_{ij}=-1$ if it leaves node $j$, and $b_{ij}=0$ otherwise,  $i=l_1,l_2,\ldots,l_{N_{\mathcal{E}}}$.

The primary cooperative objective is to achieve 
\begin{equation}\label{eq:goal}
\lim _{t\rightarrow\infty} \|x_i(t)-x_j(t)-d_{ij}\|\rightarrow 0,\, \forall \, (j,i)\in\mathcal{E},
\end{equation}
where $d_{ij}\in \mathbb{R}^n$ is the desired relative state, and let $d_{ij}=-d_{ji}$ if $(j,i)\in\mathcal{E}$. The following assumption and lemma related to the communication graph are introduced and will be employed later.
\begin{assumption}\label{ass1}
    The graph $\mathcal{G}(\mathcal{V} ,\mathcal{E},\mathcal{A} )$ is a tree.
\end{assumption}
\begin{lemma}\citep{dimarogonas2010stability}\label{le1}
   If graph $\mathcal{G}$ is a tree, then incidence matrix $B$ is full row rank.
\end{lemma}

\subsection{Communication Noise Model}
In practical MAS, information exchange between agents is inevitably affected by channel noise. To capture this, the communication system is modeled as a linear {Single-Input Single-Output (SISO)} link with additive white Gaussian noise (AWGN) for block-transmitted vector signals:
\begin{equation}\label{e2}
    \hat{y}(t)=h(t)y(t)+\eta_0(t).
\end{equation}
where $y(t)\in \mathbb{S}^n$ is the transmitted signal, $\hat{y}(t)\in \mathbb{S}^n$ is the received signal, $h(t)\in \mathbb{S}$ is the channel gain, and $\eta_0(t)$ is the aggregate additive white gaussian noise (AWGN) in the channel.


The following assumption is imposed on the communication channel.
\begin{assumption}[Channel model]\label{ass2}
The communication link is a large-scale fading AWGN channel, and the receiver is assumed to have perfect channel state information (CSI). The vector signal is transmitted in a block over orthogonal resources, and the impact of shadowing is assumed to be negligible.
\end{assumption}
Under Assumption \ref{ass2}, the channel gain $h(t)$ is real, deterministic and $h(t)\ge0$ since shadowing is negligible within the block and the availability of perfect CSI. Owing to the large-scale fading characteristics, $h(t)$ is primarily dominated by the propagation distance. The shadowing assumption is generally valid when signal propagation is not severely obstructed, and is used here for analytical simplicity rather than {being} a strict condition. The CSI assumption is commonly adopted to simplify the analysis of communication noise \citep{sery2020analog,cao2020optimized}, and the SISO large-scale fading channel model is widely considered in robotic networks and vehicular systems \citep{khuwaja2018survey}. Orthogonal resources are utilized to guarantee the orthogonality of transmitted signals, thereby preventing inter-agent interference.

After signal modulation and coherent equalization, the received signal based on Assumption \ref{ass2} becomes
\begin{equation}\label{e2}
    \hat{y}'(t)=y(t)+h^{-1}(t)\eta(t).
\end{equation}
with $\hat{y}'(t)= h^{-1}(t)\hat{y}(t)$ and $h(t)>0$. Packet loss ($h =0$) is not considered here. The model \eqref{e2} is appropriate when \textcolor{black}{the} dimension $n$ is below the order of tens of thousands \citep{liu2020privacy}.

Building on the above channel model, the received signal of agent 
$i$ from agent $j$ in the MAS described by \eqref{e1} can be expressed as
\begin{equation}\label{e3}
    \hat{x}_{ij}(t)=x_j(t)+h_{ij}^{-1}(t)z_{ij}(t)+z_i(t),
\end{equation}
where $x_j$ is the state of agent $j$, $\hat{x}_{ij}$ is the compensated received signal at agent $i$. Here, $z_{ij}\sim\mathcal{N}(0,k_{ij,1}\mathbf{I})$ with $k_{ij,1} >0$, is the \textcolor{black}{independent and identically distributed (i.i.d.)} channel noise between agents $i$ and $j$, while $z_i\sim\mathcal{N}(0,r_{i}\mathbf{I})$ with $r_{i}>0$, represents the i.i.d. receiver noise at agent $i$. The latter accounts for additive disturbances introduced after equalization (e.g., \textcolor{black}{analog-to-digital converter} quantization or finite-precision rounding) and is therefore independent of the link $(i,j)$. The random variables $z_{ij}$ and $z_i$ are mutually independent. The channel coefficient is modeled as $h_{ij}=\frac{k_{ij,2}}{\|x_i-x_j\|^{\alpha}}\mathbf{I}$ with $k_{ij,2}>0$, where $\alpha\ge 1$ quantifies how the relative distance $\|x_i-x_j\|$ affects the signal-to-noise ratio (SNR). Specifically, a greater inter-agent distance yields a smaller SNR, which captures the effect of distance-dependent path loss.

In this paper, to simplify the analysis, \textcolor{black}{$\alpha$ is set to $1$,} which is typical when $\|x_i-x_j\|$ represents the Euclidean distance between agents \citep[Chapter 4]{rappaport2010wireless}. The reception model is then simplified as
\begin{align}\label{e4}
    \hat{x}_{ij}(t)&=x_j(t)+\eta_{ij}(t),\nonumber \\
    \eta_{ij}(t)& \sim \mathcal{N}(0,(\sigma_{ij}\|x_i(t)-x_j(t)\|^2+r_{i})\mathbf{I})
\end{align}
where $\sigma_{ij}=\frac{k_{ij,1}}{k_{ij,2}^2}>0$. A similar noise model can also be found in entirely different settings, such as distributed averaging \citep{li2018distributed} and Time Difference of Arrival based localization \citep{huang2014tdoa}.

\subsection{Differential Privacy}

\textcolor{black}{The} goal is to protect reference signals $d_{ij},\, (j,i)=l_1,l_2,\ldots,l_{N_\mathcal{E}}$, which are crucial for privacy. For instance, when $d_{ij}$ represents the target relative position, its exposure may reveal the geometric configuration of a UAV formation or a sensor network localization system \citep{oh2015survey,xu2024multi}, and can even compromise the absolute positions of individual agents.

\textcolor{black}{The definition of differential privacy employed to protect $d_{ij},\, (j,i)=l_1,l_2,\ldots,l_{N_\mathcal{E}}$ is now introduced}. Define ${D}\triangleq\{d_{ij}\,|\,(j,i)=l_1,l_2,\ldots,l_{N_\mathcal{E}}\}$ as a private database to be protected, and $\mathcal{D}$ as the universe of all possible databases.  

\begin{definition}[Adjacent Relation]\label{de1}
    Two databases ${D}=\{d_{ij}\,|\,(j,i)=l_1,l_2,\ldots,l_{N_\mathcal{E}}\}$ and ${D'}=\{d'_{ij}\,|\,(j,i)=l_1,l_2,\ldots,l_{N_\mathcal{E}}\}$ are said to be adjacent if there exists a $l_k\in\{l_1,\ldots,l_{N_\mathcal{E}}\}$
    \begin{equation}\label{e5}
        \left\{
        \begin{aligned}
            &d_{ij}=d'_{ij},\, (i,j)\neq l_k, \\
            &\|d_{ij}-d'_{ij}\|_1 \leq \theta,\, (i,j)=l_k,
        \end{aligned}
        \right.
    \end{equation}
\end{definition}

A randomized mechanism $\mathcal{M}$ acting on a database $D$ is said to be differentially private if it ensures that adjacent databases $D$ and $D'$ produce nearly indistinguishable outputs $\mathcal{M}(D)$ and $\mathcal{M}(D')$.
\begin{definition}\citep{dwork2014algorithmic}\label{de2}
Let $\epsilon>0$ and $\delta>0$. A randomized algorithm $\mathcal{M}$ with $\mathrm{Range}(\mathcal{M}) \subseteq \mathbb{R}^n$ is $(\epsilon,\delta)$-differentially private if for all $S \subseteq \mathrm{Range}(\mathcal{M})$ and for all $D,D' \in \mathcal{D}$ satisfying Definition \ref{de1}, there holds
\begin{equation}
    \Pr[\mathcal{M}(D) \in S] \le \mathbf{e}^\epsilon \Pr[\mathcal{M}(D') \in S] + \delta.
\end{equation}
\end{definition}

\begin{lemma}\citep{le2014differentially}\label{le2}
 Let $\epsilon >0$, $\delta\in (0,\frac{1}{2})$ and $\mathcal{Q}(x)=\frac{1}{\sqrt{2\pi}}\int_x^\infty\mathrm{e}^{-\frac{u^2}{2}}du$. Consider a mechanism $M$ with $\ell_2$-sensitivity,
    \begin{align*}
        \Delta_{2,M}=\max\limits_{D,D'} \|M(D)-M(D')\|,
    \end{align*}
    where databases $D$ and $D'$ satisfy Definition \ref{de1}. Then the mechanism $\mathcal{M}(D)=M(D)+\eta$ is $(\epsilon,\delta)$-differentially private if each component of $\eta$ are i.i.d. and scaled to $\mathcal{N}(0,\sigma^2)$, where $\sigma \ge \frac{\Delta_{2,M}}{\sqrt{{\mathcal{Q}^{-2}(\delta)}+2\epsilon}-\mathcal{Q}^{-1}(\delta)}$.
\end{lemma}

\subsection{Problem Statement} 
Due to the stochasticity introduced for privacy preservation, the control objective in~\eqref{eq:goal} cannot be achieved exactly. Accordingly, the cooperative control problem is reformulated and formally stated as follows.
\begin{prob}\label{pr1}
Design a \textit{differentially private distributed cooperative control mechanism} for the MAS described by~\eqref{e1} such that:
\begin{enumerate}
    \item The \textcolor{black}{tracking error} $\xi_{l_k}(t)\triangleq x_i(t)-x_j(t)-d_{l_k}$ converges to a finite random limit $x_{l_k}^\ast$ in mean square and almost surely, for all $l_k=(i,j)\in\mathcal{E}$;
    \item The mechanism guarantees bounded $(\epsilon,\delta)$-differential privacy from time $t=0$ to infinity for all $d_{l_k}$, $l_k=(i,j)\in\mathcal{E}$ under the adjacency relation defined in Definition~\ref{de1}.
\end{enumerate}
\end{prob}

The first objective ensures the convergence of the \textcolor{black}{tracking error} in a stochastic sense, while the second enforces privacy protection on the reference signals. 
To facilitate the subsequent convergence analysis in  Section~\ref{subsec:Convergence}, the following result from martingale theory \textcolor{black}{is needed}.
\begin{lemma}\citep{ash2014real}\label{le3}
    Let $\{X(k),\mathcal{F}(k)\}$ be a martingale sequence satisfying $\sup\limits_{t\geq0}\mathbb{E}\|X(k)\|^2\leq \infty$. Then $X(k)$ converges almost surely and in mean square.
\end{lemma}

\section{Main Results}\label{sec:main}
In this section, a differentially private distributed finite-horizon LQR framework \textcolor{black}{is developed} to solve Problem~\ref{pr1}. The proposed approach ensures both cooperative convergence and privacy preservation under inherent communication noise. Rigorous analyzes of convergence and privacy properties are subsequently provided.

\subsection{Differentially Private Cooperative Control Algorithm}
Consider a finite horizon $T$. The finite-horizon LQR controller for each agent is obtained by recursively solving the following optimisation problem.
\begin{subequations}\label{eq:algorithm}
\begin{align}
\min\limits_{u_i(\cdot)}\quad&  J_i (x_i,\hat{x}_{ij},u_i)\\
\textbf{s.t. } &  \eqref{e1}
\end{align}
\end{subequations}
where the local objective function is defined as
\begin{align}\label{e7}
    &J_i(x(t),c(t))\triangleq \sum\limits_{k=0}^{T-1}\Biggl(u_i^{\top}(t\!+\!k)R_iu_i(t\!+\!k)+c(t)\sum\limits_{j\in \mathcal{N}_i}a_{ij}\Biggr.\nonumber\\
    \Biggl.&\quad  (x_i(t+k)-\hat{x}_{ij}(t)-d_{ij})^{\top}Q_i(x_i(t+k)-\hat{x}_{ij}(t)-d_{ij})\Biggr),
\end{align}
with $c(t)>0$ a tunable weighting sequence, $\hat{x}_{ij}(t)$ the signal received by agent $i$ according to the channel model \eqref{e4}, and $d_{ij}$ the desired relative state between agents $i$ and $j$. The matrices $Q_i \in \mathcal{Q}_i \subset {\mathbb{R}_{>0}^{n\times n}}$ and $R_i\in \mathcal{R}_i \subset {\mathbb{R}_{>0}^{n\times n}}$ are weights that balance the trade-off between cooperative accuracy and control effort.

At each time step, only the first element of the optimal control sequence is implemented. The overall procedure is summarized \textcolor{black}{in Algorithm 1}.

\begin{algorithm}
\caption{Differentially Private Cooperative Control Algorithm with Inherent Communication Noise}
\begin{algorithmic}[1]
\STATE \textbf{Initialization:} $x_i(0)\in \mathbb{R}^n$, $Q_i \in \mathcal{Q}_i$, $R_i\in \mathcal{R}_i$, $i=1,2,\ldots,N$, and $d_{ij}$, $(i,j)=l_1,l_2,\ldots,l_{N_\mathcal{E}}$.
\STATE \textbf{Set} {time series} $c(t)$ for all agents $i=1,2,\ldots,N$.
\FOR{$t=0, 1, 2, \ldots$}
    \FOR{$i=1, 2, \ldots, N$}
    \STATE Agent $i$ receives $\hat{x}_{ij}(t)$ in \eqref{e4} from its neighbour.
    \ENDFOR \ $i=N$
    \STATE Each agent $i$ obtains $u_i(t)$ by minimizing the $J_i$ in \eqref{e7} and updates the state $x_i(t+1)$ by \eqref{e1}. 
    \ENDFOR
\end{algorithmic}
\label{alg:dpcc}
\end{algorithm}

In this framework, Algorithm~\ref{alg:dpcc} can be viewed as a sequence of LQR mechanisms 
$\mathbf{\mathcal{M}}=\{\mathbf{\mathcal{M}}_0,\mathbf{\mathcal{M}}_1,\mathbf{\mathcal{M}}_2,\ldots\}$,
where $\mathbf{\mathcal{M}}_t$ represents the execution of Steps~4–7 at time~$t$. The selection rule and analytical properties of $c(t)$ will be discussed in the subsequent analysis.

\subsection{Convergence Analysis}\label{subsec:Convergence}
The unconstrained control input $u_i(t)$ in \eqref{eq:algorithm}, obtained via explicit MPC \citep{bemporad2002model}, can be expressed as
\begin{align}\label{e8}
    u_i(t) = c(t)K_{i,t}\sum\limits_{j\in\mathcal{N}_i}a_{ij}  (\hat{x}_{ij}(t)+d_{ij}-x_i(t)),
\end{align}where $K_{i,t}$ is a time-varying and positive definite gain matrix depending on $c(t)$, $Q_i$ and $R_i$. One can refer to \cite{bemporad2002model} to compute $K_{i,t}$. 
If $\mathcal{Q}_i$ and $\mathcal{R}_i$ are bounded, $K_{i,t}$ remains bounded for all $t\in\mathbb{N}$.
The closed-loop system dynamics for agent $i$ under \eqref{e8} are given by
\begin{equation}\label{e9}
    x_i(t+1)\!=\!x_i(t)+c(t)\!\sum\limits_{j\in \mathcal{N}_i}\!a_{ij}K_{i,t}(\hat{x}_{ij}(t)+d_{ij}-x_i(t)).
    \end{equation}
    
Stack all edge-related errors $\xi_{l_k}$ in Problem \ref{pr1} as $\xi=[\xi^\top_{l_1},\xi^\top_{l_2},\ldots,\xi^\top_{l_{N_\mathcal{E}}}]^\top$. 
Analogously, stack $\eta_{l_k}$ as $\eta=[\eta^\top_{l_1},\eta^\top_{l_2},\ldots,\allowbreak \eta^\top_{l_{N_\mathcal{E}}}]^\top$. The convergence properties of $\xi(t)$ are given below.

\begin{theorem}\label{th1}
Consider the MAS \eqref{e1} under Assumptions~\ref{ass1} and~\ref{ass2}. 
If the sequence $c(t)>0$ is designed such that $c(t) \in \ell_2$, 
then under Algorithm~\ref{alg:dpcc}, the \textcolor{black}{tracking error} $\xi(t)$ converges almost surely and in mean square.
\end{theorem}

\begin{pf}
Based on \eqref{e4} and \eqref{e9}, the dynamics of $\xi(t)$ are given as
\begin{equation}\label{e10}
    \xi(t+1)\!=\!\xi(t)-c(t)(B\otimes \mathbf{I}_n)K_t(B^\top W\otimes\mathbf{I}_n)[\xi(t)-\eta(t)],
\end{equation}
where $B$ is the incidence matrix, $K_t=\mathrm{diag}\{K_{1,t},\ldots,K_{N,t}\}$, and $W$ is the edge weight matrix. Define $\Psi_t\triangleq(B\otimes \mathbf{I}_n)K_t(B^\top W\otimes\mathbf{I}_n)$. Due to the boundedness of $K_{i,t},\, i=1,\ldots,N,\, \forall t \in\mathbb{N}$, there exists a $\rho_\Psi>0$ such that $\|\Psi_t\|\leq \rho_\Psi,\,\forall t \in\mathbb{N}$. 
Given the positive definiteness of $K_t$, $B$ is full row rank according to Lemma \ref{le1}, and $W=\mathbf{I}_{N_\mathcal{E}}$, $\Psi_t$ can be verified to be positive definite.

\textcolor{black}{The convergence of $\xi(t+1)$ is now analysed}. Following from \eqref{e10},
\begin{align}\label{e11}
       \xi(t+1)&=(\mathbf{I}_{n\cdot N_\mathcal{E}}-c(t)\Psi_t)\xi(t)+c(t)\Psi_t\eta(t)\nonumber\\
       &\!=\!\Phi(t,0)\xi(0)\!+\!\sum\limits_{k=0}^t\Phi(t,k+1)c(k)\Psi_k\eta(k),
\end{align}
where $\Phi(t,k)=\prod\limits_ {j=k} ^t(\mathbf{I}_{n\cdot N_\mathcal{E}}-c(j)\Psi_j)$ when $t>k\ge0$, and $\Phi(k,k)=\mathbf{I}$. The first item of \eqref{e11} is deterministic, whereas the second term 
\[
M(t)=\sum_{k=0}^{t}\Phi(t,k+1)c(k)\Psi_k\eta(k)
\]
is random and can be shown to form a martingale with respect to the 
$\sigma$-filtration $\mathcal{F}_t=\sigma(\eta(0),\ldots,\eta(t))$. 
Specifically, since each $\eta(k)$ is zero-mean and independent of 
$\mathcal{F}_{k-1}$, it follows that 
$\mathbb{E}[M(t+1)\mid\mathcal{F}_t]=M(t)$. 
Moreover, $c(k)$, $\Psi_k$, and $\Phi(t,k+1)$ are deterministic and uniformly bounded, 
ensuring that $\{M(t),\mathcal{F}_t\}$ is a martingale sequence.
\textcolor{black}{Its boundedness is now evaluated} as follows
\begin{align}\label{e12}
    &\sup\limits_{t\geq0}\mathbb{E}[\|M(t)\|^2] \nonumber\\
    &=\sup\limits_{t\geq0}\sum\limits_{k=0}^t\mathbb{E}[c^2(k)\eta^\top(k)\Psi^\top_k\Phi^\top(t,k+1)\Phi(t,k+1)\Psi_k\eta(k)] \nonumber\\
    &\leq \sup\limits_{t\geq0} \sum\limits_{k=0}^t\rho^2_\Psi c^2(k)\|\Phi(t,k+1)\|^2\mathbb{E}[\eta^\top(k)\eta(k)].
\end{align}
With $c(t)\rightarrow0$ as $t\rightarrow\infty$ and $\Psi_t \succ 0$, there exists a $T_c>0$ such that $\mathbf{I}-c(t)\Psi_t\succ 0$ and $\|\mathbf{I}-c(t)\Psi_t\|<1,\,\forall\, t\geq T_c$. It then follows that $
    \|\Phi(\infty,T_c)\|^2\leq \prod_ {j=T_c} ^\infty\|\mathbf{I}_{nN_\mathcal{E}}-c(j)\Psi_j\|^2<1$, which implies the existence of $\rho_\Phi>0$ such that $\|\Phi(t,k)\|^2\leq\rho_\Phi$ for all $t\ge k$. Based on \eqref{e4},
    \begin{align}\label{e13}
        \mathbb{E}[\eta^\top(t)\eta(t)]&=\mathbb{E}[\sum\limits_{k={l_1}}^{l_{N_\mathcal{E}}}\eta_k^\top(t)\eta_k(t)]\nonumber\\
        &= n\sum\limits_{k={l_1}}^{l_{N_\mathcal{E}}}\mathbb{E}[\sigma_{k}\|\xi_k(t)-d_k\|^2+r_{k}]\nonumber\\
        &\leq n\sum\limits_{k={l_1}}^{l_{N_\mathcal{E}}}\mathbb{E}[2\sigma_{k}\|\xi_k(t)\|^2+2\sigma_{k}\|d_k\|^2+r_{k}]\nonumber\\
        &\leq 2n\sigma d^\top d+nrN_\mathcal{E}+2n\sigma\mathbb{E}[\xi^\top(t)\xi(t)]
    \end{align}
 where $\sigma=\max\{\sigma_{l_1},\ldots,\sigma_{l_{N_\mathcal{E}}}\}$, $r=\max\{r_{1},\ldots,r_{N}\}$, and $d=[d^\top_{l_1},\ldots,d^\top_{l_{N_\mathcal{E}}}]^\top$. Following from \eqref{e10},
 \begin{align*}
\xi^\top(t+1)&\xi(t+1)=\xi^\top(t)\xi(t)-2c(t)\xi^\top(t)\Psi_t\xi(t)\\
&+c^2(t)\xi^\top(t)\Psi_t^\top\Psi_t\xi(t)+c^2(t)\eta^\top(t)\Psi_t^\top\Psi_t\eta(t)\\
     &+\xi^\top(t)[c(t)\Psi_t-c^2(t)\Psi_t^\top\Psi_t]\eta(t)\\
     &+\eta^\top(t)[c(t)\Psi_t-c^2(t)\Psi_t^\top\Psi_t]\xi(t).
 \end{align*}
By taking expectations on both sides, the two cross terms vanish, i.e.,
$
\mathbb{E}\!\left[\xi^\top(t)\!\left(c(t)\Psi_t - c^2(t)\Psi_t^\top\Psi_t\right)\eta(t)\right] = 0,$
$\mathbb{E}\!\left[\eta^\top(t)\!\left(c(t)\Psi_t - c^2(t)\Psi_t^\top\Psi_t\right)\xi(t)\right] = 0,
$
owing to the independence between $\eta(t)$ and $\xi(t)$ and the zero-mean property $\mathbb{E}[\eta(t)] = 0$. 
Consequently,
 \begin{multline}\label{e14}
     \mathbb{E}[\xi^\top(t+1)\xi(t+1)]=\mathbb{E}[\xi^\top(t)\xi(t)]-2c(t)\mathbb{E}[\xi^\top(t)\Psi_t\xi(t)]\\
     +c^2(t)\mathbb{E}[\xi^\top(t)\Psi_t^\top\Psi_t\xi(t)]+c^2(t)\mathbb{E}[\eta^\top(t)\Psi_t^\top\Psi_t\eta(t)]\\
     \leq \!(1\!-\!2c(t)\rho_{\Psi}\!+\!c^2(t)\rho_{\Psi}^2)\mathbb{E}[\xi^\top(t)\xi(t)]\!+\!c^2(t)\rho_{\Psi}^2\mathbb{E}[\eta^\top(t)\eta(t)].
 \end{multline}
Substituting \eqref{e13} into \eqref{e14},
\begin{equation}\label{e15}
    \begin{aligned}
        &\mathbb{E}[\xi^\top(t\!+\!1)\xi(t\!+\!1)]\!\leq\!(1\!-\!2c(t)\rho_{\Psi}\!+\!c^2(t)\rho_{\Psi}^2
    +2n\sigma c^2(t)\rho_{\Psi}^2)\\ &\times \mathbb{E}[\xi^\top(t)\xi(t)]+2n\sigma c^2(t)\rho_{\Psi}^2d^\top d+nrN_\mathcal{E}c^2(t)\rho_{\Psi}^2\\ &\leq\! \phi(t,0)
\!\!\times\!\!\xi^\top\!(0)\xi(0)\!\!+\!\!\sum\limits_{k=0}^t\phi(t,k\!+\!1)c^2(k)\rho_{\Psi}^2[2n\sigma d^\top d\!\!+\!\!nrN_\mathcal{E}],
    \end{aligned}
\end{equation}
where $\phi(t,k)=\prod\limits_ {j=k} ^t(1\!-\!2c(j)\rho_{\Psi}\!+\!c^2(j)\rho_{\Psi}^2
    +2n\sigma c^2(j)\rho_{\Psi}^2)$, when $t>k\ge0$, and $\phi(k,k)=1$. 
Applying the inequality $1+w \le e^{w},\,\forall w\in\mathbb{R}$,
\begin{multline*}
\phi(t,k)
=\prod_{j=k}^{t}\!\left(1 - 2c(j)\rho_{\Psi} + c^2(j)\rho_{\Psi}^2 + 2n\sigma c^2(j)\rho_{\Psi}^2\right)
\\
\leq
\exp\!\Bigg(\sum_{j=k}^{t}\!\!\left[-2c(j)\rho_{\Psi} + c^2(j)\rho_{\Psi}^2 + 2n\sigma c^2(j)\rho_{\Psi}^2\right]\!\Bigg).
\end{multline*}
Since $c(t)\!\in\!\ell_2$, it follows that $\sum_{j=0}^{\infty} c^2(j)\!<\!\infty$ and $c(t)\!\to\!0$, 
which ensures that the exponent is bounded from above. 
Hence, there exists a constant $\rho_{\phi}>0$ such that
\[
\phi(t,k)\le \rho_{\phi},\quad \forall\, t\ge k.
\]
Combining \eqref{e12}-\eqref{e15} yields $ \sup\limits_{t\geq0}\mathbb{E}[\|M(t)\|^2]< \infty$. By Lemma \ref{le3}, $\{M(t),\mathcal{F}_t\}$ is thus a square-integrable martingale 
and therefore converges almost surely and in mean square to a finite random limit $M^*$. Consequently, from \eqref{e14},
\begin{equation}\label{19}
\xi(t)\to \xi^* := \Phi(\infty,0)\,\xi(0) + M^*, \quad \text{a.s.\ and in mean square,}
\end{equation}
where $\Phi(\infty,0)$ exists 
since $\prod_{j=0}^\infty (I - c(j)\Psi_j)$ converges under $c(j)\in\ell_2$ and $\Psi_j\succ0$.
\end{pf}

\begin{remark}
\textcolor{black}{By taking the expectation on the right side of equation \eqref{19} and referring to \eqref{e12},} it can be verified that under the Assumptions \ref{ass1} and \ref{ass2}, $\mathbb{E}[\xi^\top(t)\xi(t)]<\infty,\, \forall t\ge 0$, $\|\mathbb{E}[\xi^\ast]\|<\|\xi(0)\|$ and $\mathrm{var}(\xi^\ast)<\infty$. Furthermore, if the design parameter $c(t)\in\ell_2\setminus \ell_1$, then $\mathbb{E}[\xi^\ast]=0$. The detailed derivation is omitted here due to space limitations and will be presented in our forthcoming article. Based on the boundedness of $\mathbb{E}[\xi^\top(t)\xi(t)]$ and the zero-mean property of the communication noise, it follows from \eqref{e13} that the variance of the noise remains bounded for all $t$.
\end{remark}

\subsection{Privacy Analysis}
In this subsection, the privacy-preserving performance of Algorithm 1 \textcolor{black}{is analysed}. 
{\color{black}To distinguish from $(\epsilon, \delta)$ privacy defined over the infinite horizon, let us introduce $(\epsilon_t, \delta_t)$ as the instantaneous privacy parameters at step $t$. Here, $\delta_t$ quantifies the probability that differential privacy fails and is chosen to balance the trade-off between privacy protection and control performance.}
\begin{theorem}
Suppose Assumptions \ref{ass1} and \ref{ass2} hold, and $c(t)$ and $\delta_t$ are designed such that
\begin{equation}\label{con1}
c(t)>0,\,\,\, c(t) \in \ell_2, \,\,\delta_t \in \ell_1,\,\,  c(t)\mathcal{Q}^{-1}(\delta_t)\in \ell_1\,.
\end{equation}
Algorithm 1 can ensure bounded $(\epsilon,\delta)$-DP even as the number of iterations approaches infinity.
\end{theorem}
\begin{pf}
\textcolor{black}{The privacy property of the LQR mechanism $\mathcal{M}_t$ at time $t$ is first analyzed}. Let $D$ and $D'$ be two adjacent databases as defined in Definition \ref{de1}. The corresponding state vectors generated by the mechanism are {$\mathbf{\mathcal{M}}_t(D)=\left[
        x_1^{\top}(t+1),  x_2^{\top}(t+1), \cdots, x^{\top}_N(t+1)
\right]^{\top}$ and $\mathbf{\mathcal{M}}_t(D')=\left[
        x'^{\top}_1(t+1) , x'^{\top}_2(t+1) , \cdots, x'^{\top}_N(t+1)
\right]^{\top}$.}
For any pair of adjacent databases, only the states of the two agents corresponding to the differing edge $(i,j)=l_k$ are affected, i.e., 
$x_q(t\!+\!1)=x'_q(t\!+\!1)$ for all $q\neq i,j$.  
Hence, the $\ell_2$-sensitivity of $\mathcal{M}_t$ is
\begin{align*}
    \Delta_{2,M,t}&=\max\limits_{D,D'} \|M_t(D)-M_t(D')\|\\
   &=\max\limits_{D,D'} \sqrt{\|x_{i,t+1}-x_{i,t+1}'\|^2+\|x_{j,t+1}-x_{i,t+1}'\|^2}\\
   &\leq 2c(t)\rho_{K,t}\theta
\end{align*}
where $\rho_{K,t}=\max\{\|K_{1,t}\|,\ldots,\|K_{N,t}\|\}$ and $\theta$ is the adjacency bound defined in~\eqref{e5}. Note that based on Lemma \ref{le2}, the {\color{black} instantaneous} privacy budget $\epsilon_t$ of $\mathcal{M}_t$ given the channel noise in \eqref{e4} satisfies 
\begin{align}
    \epsilon_t&\leq \max\limits_{(i,j)\in \mathcal{N}}\frac{\Delta_{2,M,t}^2}{2(\sigma_{ij}\|x_i(t)-x_j(t)\|^2+r_{i})}\nonumber\\&+\frac{\Delta_{2,M,t}}{\sqrt{\sigma_{ij}\|x_i(t)-x_j(t)\|^2+r_{i}}}\mathcal{Q}^{-1}(\delta_t)\nonumber\\
    &\leq \frac{2c^2(t)\rho^2_{K}\theta^2}{\underline{r}}+\frac{2c(t)\rho_{K}\theta_t}{\sqrt{\underline{r}}}\mathcal{Q}^{-1}(\delta_t)
\end{align}
where $\underline{r}=\min\{r_{1},\ldots,r_{N}\}$, $\rho_K\!\ge\!\|K_{i,t}\|$ for all $i$ and $t$. 
Thus, $\mathcal{M}_t$ preserves at least $(\epsilon_t,\delta_t)$-DP at each iteration~$t$.

Since the range of $\mathcal{M}_t(D)$ depends adaptively on previous mechanisms $\mathcal{M}_0(D),\mathcal{M}_1(D),\ldots,\mathcal{M}_{t-1}(D)$, the Adaptive Sequential Composition Theorem~\citep{dwork2014algorithmic} gives $\delta=\sum_{t=0}^\infty \delta_t<\infty$ due to $\delta_t\in\ell_1$, and
\begin{align*}
    \epsilon=\sum_{t=0}^\infty \epsilon_t\leq \frac{2\rho^2_{K}\theta^2}{\underline{r}}\sum\limits_{t=0}^\infty c^2(t)+\frac{2\rho_{K}\theta_t}{\sqrt{\underline{r}}}\sum\limits_{t=0}^\infty c(t)\mathcal{Q}^{-1}(\delta_t).
\end{align*} 
Because $c(t)\in\ell_2$ and $ c(t)\mathcal{Q}^{-1}(\delta_t)\in \ell_1$, it is immediate to show $\epsilon<\infty.$ Therefore, Algorithm~\ref{alg:dpcc} ensures bounded $(\epsilon,\delta)$-differential privacy as the number of iterations tends to infinity. The proof is complete.
\end{pf}
\begin{remark}
Since $\mathcal{Q}^{-1}(\delta_t)=O\Big(\sqrt{2 \ln \frac{1}{\delta_t}}\Big)$, this observation can help simplify the selection of $c(t)$ and $\delta_t$ in \eqref{con1}. In fact, equation \eqref{con1} is quite conservative. This conservatism arises not only from certain relaxations introduced during the proof, but also from the stringent requirement of maintaining bounded privacy over an infinite number of iterations. In practice, this would correspond to an eavesdropper monitoring the system from time zero to infinity, which is generally unrealistic.
\end{remark}


\subsection{Numerical Simulation}
\begin{figure}[htp!]
 \centering
  \includegraphics[width=0.5\textwidth]{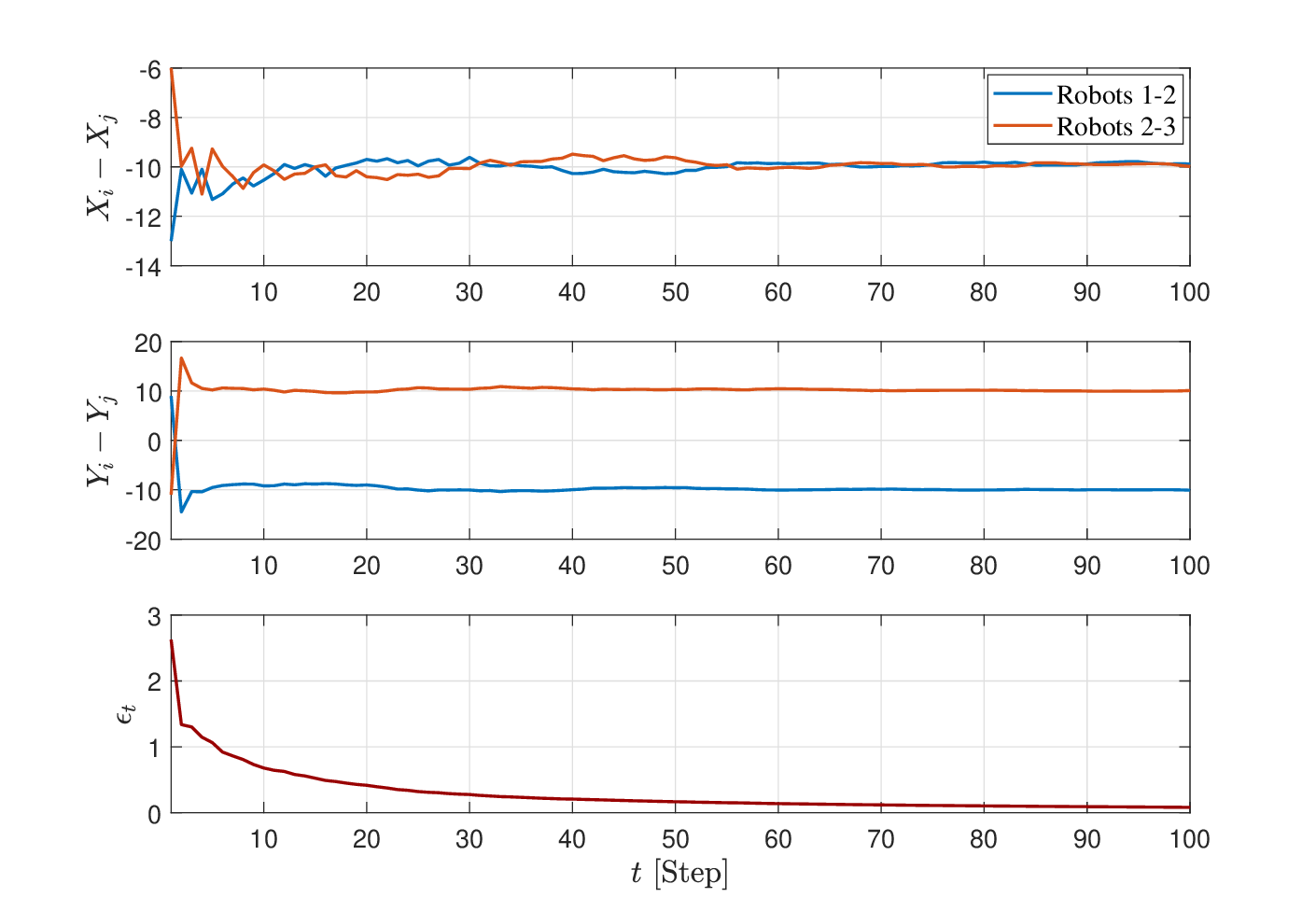}
  \caption{\centering Evolution of Relative States and Privacy Budget}
  \label{fig.1}
\end{figure}

\begin{figure}[htp!]
 \centering
  \includegraphics[width=0.5\textwidth]{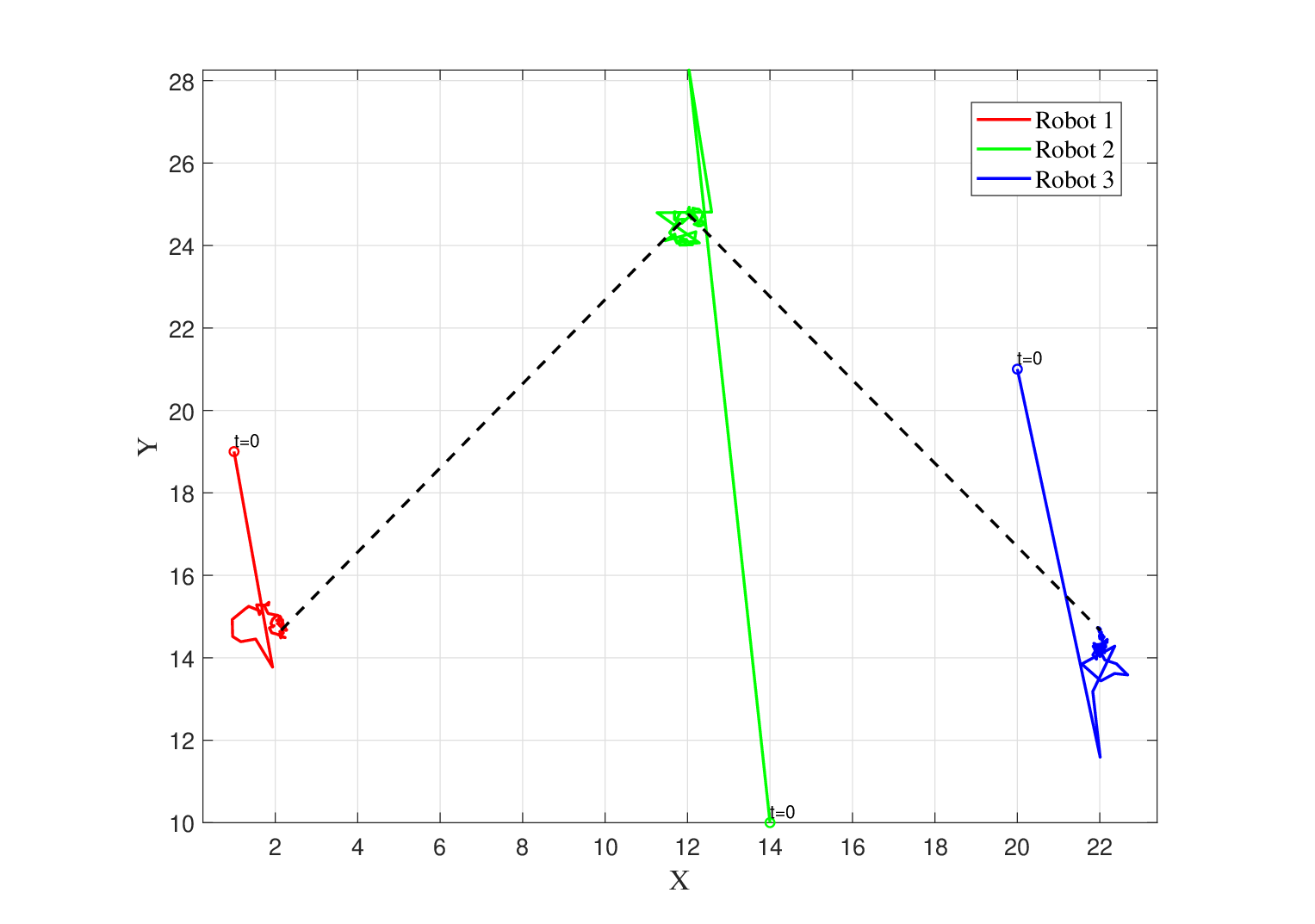}
  \caption{\centering Trajectories of Robots}
  \label{fig.2}
\end{figure}

Consider a $3$-robot system operating in a $2$-dimensional space. The dynamics of each robot are described by \eqref{e1}, where the state represents the $(x,y)$-position coordinates of the robot. The undirected communication graph is given by 
\begin{align*}
    \mathcal{A}=\left[\begin{matrix}
        0 & 1 & 0 \\1 & 0 &1 \\0 & 1 &0
    \end{matrix}\right],\quad B=\left[\begin{matrix}
        1 & -1 & 0 \\0 & 1 &-1
    \end{matrix}\right].
\end{align*}The initial positions are $x_1(0)=\left[\begin{matrix}1 &19\end{matrix}\right]^{\top}$, $x_2(0)=\left[\begin{matrix}14 &10 \end{matrix}\right]^{\top}$, $x_3(0)=\left[\begin{matrix}20 &21 \end{matrix}\right]^{\top}$. The desired relative positions are $d_{12}=\left[\begin{matrix}-10 & -10\end{matrix}\right]^{\top}$, $d_{23}=\left[\begin{matrix}-10 &10\end{matrix}\right]^{\top}$. Let $\theta=1$ in \eqref{e5}. Let $\sigma_{ij}=0.01,\,r_{i}=0.1, i=1,2,3,\, \forall(i,j)\in\mathcal{E}$ in \eqref{e4}. Set the rolling horizon length and $Q_i$, $R_i$ respectively to $T=10$ and  
\begin{align*}
    Q_i=\left[\begin{matrix}
       8 &0 \\0 &8 
    \end{matrix}\right],  {R}_i=\left[\begin{matrix}
      3 &0 \\0 &3
    \end{matrix}\right],\ i=1,2,3.
\end{align*} \textcolor{black}{Select} $c(t)=\frac{1}{7(t+1)^{1.26}}$ and $\delta_t=0.001\mathbf{e}^{-\sqrt{t}}, \forall t\in\mathbb{N}$, which can be verified to satisfy \eqref{con1}. The evolution of the relative states of neighbouring robots and the algorithm's privacy budget under Algorithm 1 are illustrated in Fig.~\ref{fig.1}. The position trajectories of robots are shown in Fig.~\ref{fig.2}. When an adversary continuously monitors the communication from $t_1 = 0$ to $t_2 = 100$, the algorithm guarantees $(\epsilon, \delta)$-differential privacy with $\epsilon=\sum\limits_{t=0}^{100}\epsilon_t=29.9587$, $\delta=\sum\limits_{t=0}^{100}\delta_t=0.0017$.

\section{Conclusion}
This paper studies privacy-preserving cooperative control in multi-agent systems under the framework of differential privacy. By considering both channel and receiver noise, a realistic relative-state-dependent noise model is established. It is then shown that, under the proposed LQR-based cooperative control strategy, the inherent communication noise can naturally protect the agents’ reference signals with bounded $(\epsilon,\delta)$-differential privacy. At the same time, the algorithm ensures that the \textcolor{black}{tracking error} can converge to a finite random limit in mean square and almost surely. 

\balance
\bibliography{list}             

@article{oh2015survey,
  title={A survey of multi-agent formation control},
  author={Oh, Kwang-Kyo and Park, Myoung-Chul and Ahn, Hyo-Sung},
  journal={Automatica},
  volume={53},
  pages={424--440},
  year={2015},
  publisher={Elsevier}
}

@article{wang2022cooperative,
  title={Cooperative and competitive multi-agent systems: from optimization to games},
  author={Wang, Jianrui and Hong, Yitian and Wang, Jiali and Xu, Jiapeng and Tang, Yang and Han, Qing-Long and Kurths, J{\"u}rgen},
  journal={IEEE/CAA Journal of Automatica Sinica},
  volume={9},
  number={5},
  pages={763--783},
  year={2022},
  publisher={IEEE}
}

@article{d2025hierarchical,
  title={Hierarchical structures in multi-agent systems: containment control and opinion dynamics},
  author={D’Alfonso, Luigi and Fedele, Giuseppe},
  journal={Systems $\&$ Control Letters},
  volume={204},
  pages={106188},
  year={2025},
  publisher={Elsevier}
}

@article{li2018cooperative,
  title={Cooperative Output Regulation of Heterogeneous Linear Multi-Agent Networks via ${H}_\infty$ Performance Allocation},
  author={Li, Xianwei and Soh, Yeng Chai and Xie, Lihua and Lewis, Frank L},
  journal={IEEE Transactions on Automatic Control},
  volume={64},
  number={2},
  pages={683--696},
  year={2018},
  publisher={IEEE}
}

@article{cao2025proximal,
  title={Proximal cooperative aerial manipulation with vertically stacked drones},
  author={Cao, Huazi and Shen, Jiahao and Zhang, Yin and Fu, Zheng and Liu, Cunjia and Sun, Sihao and Zhao, Shiyu},
  journal={Nature},
  pages={1--8},
  year={2025},
  publisher={Nature Publishing Group UK London}
}

@article{wang2023tailoring,
  title={Tailoring gradient methods for differentially private distributed optimization},
  author={Wang, Yongqiang and Nedi{\'c}, Angelia},
  journal={IEEE Transactions on Automatic Control},
  volume={69},
  number={2},
  pages={872--887},
  year={2023},
  publisher={IEEE}
}

@inproceedings{dwork2006calibrating,
  title={Calibrating noise to sensitivity in private data analysis},
  author={Dwork, Cynthia and McSherry, Frank and Nissim, Kobbi and Smith, Adam},
  booktitle={Theory of Cryptography Conference},
  pages={265--284},
  year={2006},
  organization={Springer}
}

@article{dwork2014algorithmic,
  title={The algorithmic foundations of differential privacy},
  author={Dwork, Cynthia and Roth, Aaron and others},
  journal={Foundations and Trends in Theoretical Computer Science},
  volume={9},
  number={3--4},
  pages={211--407},
  year={2014},
  publisher={Now Publishers, Inc.}
}

@book{annaswamy2023cyber,
  title={Cyber-physical-human systems: fundamentals and applications},
  author={Annaswamy, Anuradha M and Khargonekar, Pramod P and Spurgeon, Sarah K and others},
  year={2023},
  publisher={John Wiley \& Sons}
}

@article{nozari2017differentially,
  title={Differentially private average consensus: Obstructions, trade-offs, and optimal algorithm design},
  author={Nozari, Erfan and Tallapragada, Pavankumar and Cort{\'e}s, Jorge},
  journal={Automatica},
  volume={81},
  pages={221--231},
  year={2017},
  publisher={Elsevier}
}

@article{nozari2015differentially,
  title={Differentially private average consensus with optimal noise selection},
  author={Nozari, Erfan and Tallapragada, Pavankumar and Cort{\'e}s, Jorge},
  journal={IFAC-PapersOnLine},
  volume={48},
  number={22},
  pages={203--208},
  year={2015},
  publisher={Elsevier}
}

@inproceedings{hawkins2020differentially,
  title={Differentially private formation control},
  author={Hawkins, Calvin and Hale, Matthew},
  booktitle={2020 59th IEEE Conference on Decision and Control (CDC)},
  pages={6260--6265},
  year={2020},
  organization={IEEE}
}

@inproceedings{cortes2016differential,
  title={Differential privacy in control and network systems},
  author={Cort{\'e}s, Jorge and Dullerud, Geir E and Han, Shuo and Le Ny, Jerome and Mitra, Sayan and Pappas, George J},
  booktitle={2016 IEEE 55th Conference on Decision and Control (CDC)},
  pages={4252--4272},
  year={2016},
  organization={IEEE}
}

@article{huang2024differential,
  title={Differential privacy in distributed optimization with gradient tracking},
  author={Huang, Lingying and Wu, Junfeng and Shi, Dawei and Dey, Subhrakanti and Shi, Ling},
  journal={IEEE Transactions on Automatic Control},
  volume={69},
  number={9},
  pages={5727--5742},
  year={2024},
  publisher={IEEE}
}

@article{liu2020privacy,
  title={Privacy for free: wireless federated learning via uncoded transmission with adaptive power control},
  author={Liu, Dongzhu and Simeone, Osvaldo},
  journal={IEEE Journal on Selected Areas in Communications},
  volume={39},
  number={1},
  pages={170--185},
  year={2020},
  publisher={IEEE}
}

@article{liu2022wireless,
  title={Wireless federated langevin monte carlo: repurposing channel noise for bayesian sampling and privacy},
  author={Liu, Dongzhu and Simeone, Osvaldo},
  journal={IEEE Transactions on Wireless Communications},
  volume={22},
  number={5},
  pages={2946--2961},
  year={2022},
  publisher={IEEE}
}

@article{liu2024differentially,
  title={Differentially private over-the-air federated learning over MIMO fading channels},
  author={Liu, Hang and Yan, Jia and Zhang, Ying-Jun Angela},
  journal={IEEE Transactions on Wireless Communications},
  volume={23},
  number={8},
  pages={8232--8247},
  year={2024},
  publisher={IEEE}
}

@article{sery2020analog,
  title={On analog gradient descent learning over multiple access fading channels},
  author={Sery, Tomer and Cohen, Kobi},
  journal={IEEE Transactions on Signal Processing},
  volume={68},
  pages={2897--2911},
  year={2020},
  publisher={IEEE}
}

@article{cao2020optimized,
  title={Optimized power control for over-the-air computation in fading channels},
  author={Cao, Xiaowen and Zhu, Guangxu and Xu, Jie and Huang, Kaibin},
  journal={IEEE Transactions on Wireless Communications},
  volume={19},
  number={11},
  pages={7498--7513},
  year={2020},
  publisher={IEEE}
}

@article{khuwaja2018survey,
  title={A survey of channel modeling for UAV communications},
  author={Khuwaja, Aziz Altaf and Chen, Yunfei and Zhao, Nan and Alouini, Mohamed-Slim and Dobbins, Paul},
  journal={IEEE Communications Surveys $\&$ Tutorials},
  volume={20},
  number={4},
  pages={2804--2821},
  year={2018},
  publisher={IEEE}
}

@article{li2018distributed,
  title={Distributed averaging with random network graphs and noises},
  author={Li, Tao and Wang, Jiexiang},
  journal={IEEE Transactions on Information Theory},
  volume={64},
  number={11},
  pages={7063--7080},
  year={2018},
  publisher={IEEE}
}

@book{rappaport2010wireless,
  title={Wireless communications: principles and practice},
  author={Rappaport, Theodore S},
  year={2010},
  publisher={Cambridge University Press}
}

@inproceedings{xu2024multi,
  title={A Multi-Player Potential Game Approach for Sensor Network Localization with Noisy Measurements},
  author={Xu, Gehui and Chen, Guanpu and Fidan, Baris and Hong, Yiguang and Qi, Hongsheng and Parisini, Thomas and Johansson, Karl H},
  booktitle={2024 IEEE 63rd Conference on Decision and Control (CDC)},
  pages={4494--4499},
  year={2024},
  organization={IEEE}
}

@article{bemporad2002model,
  title={Model predictive control based on linear programming\~{} the explicit solution},
  author={Bemporad, Alberto and Borrelli, Francesco and Morari, Manfred and others},
  journal={IEEE Transactions on Automatic Control},
  volume={47},
  number={12},
  pages={1974--1985},
  year={2002}
}

@article{le2014differentially,
  title={Differentially private filtering},
  author={Le Ny, Jerome and Pappas, George J},
  journal={IEEE Transactions on Automatic Control},
  volume={59},
  number={2},
  pages={341--354},
  year={2014},
  publisher={IEEE}
}

@article{dimarogonas2010stability,
  title={Stability analysis for multi-agent systems using the incidence matrix: quantized communication and formation control},
  author={Dimarogonas, Dimos V and Johansson, Karl H},
  journal={Automatica},
  volume={46},
  number={4},
  pages={695--700},
  year={2010},
  publisher={Elsevier}
}

@book{ash2014real,
  title={Real analysis and probability: probability and mathematical statistics: a series of monographs and textbooks},
  author={Ash, Robert B},
  year={2014},
  publisher={Academic press}
}

@inproceedings{huang2012differentially,
  title={Differentially private iterative synchronous consensus},
  author={Huang, Zhenqi and Mitra, Sayan and Dullerud, Geir},
  booktitle={Proceedings of the 2012 ACM Workshop on Privacy in the Electronic Society},
  pages={81--90},
  year={2012}
}

@article{huang2014tdoa,
  title={TDOA-based source localization with distance-dependent noises},
  author={Huang, Baoqi and Xie, Lihua and Yang, Zai},
  journal={IEEE Transactions on Wireless Communications},
  volume={14},
  number={1},
  pages={468--480},
  year={2014},
  publisher={IEEE}
}

@article{liu2026design,
  title={Design of Stochastic Quantizers for Privacy-Preserving Control},
  author={Liu, Le and Kawano, Yu and Cao, Ming},
  journal={IEEE Transactions on Automatic Control},
  year={2026},
  publisher={IEEE}
}
                                                    
\end{document}